# Suppression of twinning and enhanced electronic anisotropy of SrIrO$_3$ films

A. K. Jaiswal[1,2], R. Schneider[1], R. Singh[2], and D. Fuchs[1,*]

[1]Karlsruhe Institute of Technology, Institute for Solid-State Physics, 76021 Karlsruhe, Germany

[2]Indian Institute of Technology Delhi, Department of Physics, New Delhi 110016, India

The spin-orbit coupling and electron correlation in perovskite SrIrO$_3$ (SIO) strongly favor new quantum states and make SIO very attractive for next generation quantum information technology. In addition, the small electronic band-width offers the possibility to manipulate anisotropic electronic transport by strain. However, twinned film growth of SIO often masks electronic anisotropy which could be very useful for device applications. We demonstrate that the twinning of SIO films on (001) oriented SrTiO$_3$ (STO) substrates can be strongly reduced for thin films with thickness $t$ less than 30 nm by using substrates displaying a TiO$_2$-terminated surface with step-edge alignment parallel to the *a*- or *b*-axis direction of the substrate. This allows us to study electronic anisotropy of strained SIO films which hitherto has been reported only for bulk-like SIO. For films with $t < 30$ nm electronic anisotropy increases with increasing $t$ and becomes even twice as large compared to nearly strain-free films grown on (110) DyScO$_3$. The experiments demonstrate the high sensitivity of electronic transport towards structural distortion and the possibility to manipulate transport by substrate engineering.

The competing interaction of spin-orbit coupling, electron correlation and crystal field splitting in the 5*d* transition metal iridates [1,2] strongly favor the appearance of new topological phenomena or quantum states [3-12] and make these materials especially advantageous for new quantum devices. The metastable form of perovskite SrIrO$_3$ (SIO) can be stabilized by epitaxial growth of thin films [13-17] and is structurally compatible to many other functional oxides [18]. Therefore, SIO films are of current interest. They also may act as a key building block for engineering new topological phases and enabling the design of heterostructures for new oxide electronics.

In SIO, tilts and rotations of the IrO$_6$ octahedra result in an orthorhombic structure with space group Pbnm (62) [19]. The hybridization of the Ir5*d* and O2*p* orbitals results in a paramagnetic semimetallic ground state [20,21]. The Fermi surface consists of multiple heavy hole- and light electronlike sheets with narrow electronic bandwidths [8,21]. The 2-6 times lighter effective mass of the electrons results in a dominant electronlike single-type charge-carrier transport [17,22]. For bulk-like SIO films, a distinct temperature $T$ dependent anisotropic electronic transport with smallest resistivity along the *c*-axis direction is found [23]. The electronic anisotropy is very likely related to the structural, i. e., orthorhombic distortion of SIO. The electronic structure is indeed controlled by a subtle interplay between octahedral rotations, SOC, and dimensionality [17,21,24], which enables concrete tuneability of electronic properties by epitaxial strain and film thickness. For example, for SIO on (001) oriented SrTiO$_3$ (STO) a metal-to-insulator transition (MIT) is observed for film thickness $t \leq 3$ unit cells [24]. The charge gap opening is accompanied by a transition to a structural phase where in-plane rotations of the IrO$_6$ octahedra are suppressed. The surface symmetry of cubic STO not only triggers a MIT for $t \leq 3$ unit cells but also results in an in-plane twinning of orthorhombic bulk-like thick SIO films [23].

---

[*]e-mail corresponding author: dirk.fuchs@kit.edu

Consequently, anisotropic electronic transport is compensated to large extent. In this Letter, we demonstrate that twinning of SIO can be suppressed when films ($t < 30$ nm) are grown on $TiO_2$-terminated (001) STO with step-edge alignment parallel to the *a*- or *b*- axis of STO. In comparison to bulk-like SIO electronic anisotropy of the epitaxial films is found to be significantly enhanced by compressive strain.

Epitaxial perovskite SIO films were prepared by pulsed laser deposition. Details are described elsewhere [23]. Films were grown on standard (001) oriented STO, i. e., as delivered from the supplier (CrysTec company), $TiO_2$-terminated STO (Ti-STO) and (110) $DyScO_3$ (DSO) simultaneously. To prevent possible surface degradation or decomposition, all the films were capped with a 4 nm thick epitaxial STO capping layer. SIO films were deposited with different film thickness, $2$ nm $< t < 60$ nm. $t$ was deduced from x-ray reflectivity, see Fig. 1a, which likewise demonstrates smooth film growth with negligible interface roughness. For SIO on STO, diffraction in the vicinity of the (110) lattice plane shows small shift of the maximum peak position to slightly smaller 2θ-values, i. e., larger out-of-plane lattice spacing $d_{110}$, which increases from 3.98 Å for $t = 60$ nm to 4.06 Å for $t = 2.9$ nm. The increase of $d_{110}$ indicates increasing compressive in-plane lattice strain on SIO with decreasing film thickness. Laue oscillations evidence layered growth up to $t \geq 60$nm. Films on DSO display similar behavior, however, due to the smaller lattice mismatch (-0.22% compared to -1.36% for SIO on STO) bulk-like structural properties ($d_{110} = 3.96$ Å) are obtained for $t = 60$ nm.

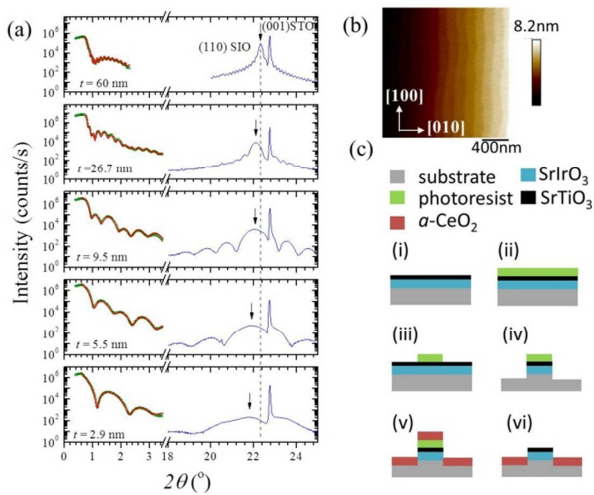

Figure 1. (a) Measured x-ray reflectivity (green line) and simulation (red line) and symmetric x-ray diffraction (blue line) for capped SIO films with different thickness on STO. Dashed line indicates position for bulk SIO and arrows the peak maximum of the film. (b) Surface topography of Ti-STO. Step-edges are parallel to the [100] direction. (c) Process scheme of photolithography: (i) film deposition by PLD, (ii) coating with a photoresist, (iii) ultra-violet light exposure and development, (iv) Ar-ion milling, (v) deposition of $CeO_2$ hard mask by PLD, and (vi) lift-off process of photoresist.

The $TiO_2$-termination process and step-edge alignment of the STO substrates were characterized by atomic force microscopy, see Fig. 1b. The steps display well defined height of one unit cell and a terrace width of about 150 nm. In the following, we only used Ti-STO substrates with step-edge alignment parallel to the [100] direction over the complete substrate surface ($5 \times 5$ mm$^2$).

Measurements of the electronic transport were carried with a physical property measuring system (PPMS) on non-patterned films by van der Pauw (VDP) technique and on microbridges (20μm ×100μm) in Hall-bar geometry by standard four-point probe (FPP) method. The microbridges were patterned along two orthogonal directions by standard photolithography and Ar-ion milling, see Fig. 1c. A 70 nm thick insulating $CeO_2$ hard mask was deposited to enable identifying Hall-bars and contact regions for very thin films. To compensate oxygen loss and small parasitic conductivity of STO after ion milling, the samples were post-annealed in flowing $O_2$ at 500°C for 5 hours.

To demonstrate suppressed twinning for SIO growth on Ti-STO we deposited films under the same conditions on STO and Ti-STO. We already documented twinned orthorhombic growth of thick SIO films on STO [23]. In order to characterize the orthorhombic structure, distortion and twinning of (110) oriented SIO films, lattice reflections with $h \neq k$ and $l = |h-k|$ have to be studied. To this end, the (444), (260), (44-4) and (620) SIO reflections were chosen. In pseudo-cubic notation, the peak family corresponds to the {204}type lattice reflections. The large 2θ values provide sufficient resolution while peak intensity is still large enough. In Fig. 2a we show the reciprocal space map (RSM) of SIO ($t$ = 60 nm) in the vicinity of the (204) STO substrate reflection for the four different azimuth directions. Since twinning of SIO results from the orthorhombic distortion, the orthorhombic notation is used in the following to characterize structural properties of SIO. Corresponding orthorhombic lattice reflections appearing in the vicinity of the cubic {204} lattice reflections of STO are the (444), (260), (44-4) and (620) SIO reflections. The twinned growth of SIO results in a peak broadening along the $l$-direction. In addition, lattice relaxation significantly smears out peak intensity towards smaller $h$- and larger $l$-values, i. e., larger in-plane and smaller out-of plane lattice parameters. Consequently, all the four film peaks appear at similar ($l,h$) position, masking the orthorhombic symmetry.

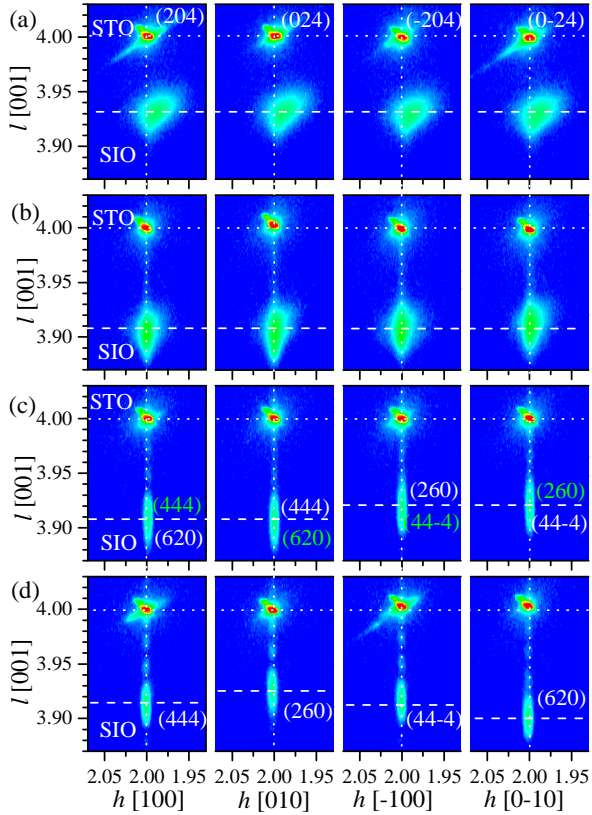

Figure 2. Contour plots displaying reciprocal space maps of SIO ($t$ = 60 nm) on STO (a) and Ti-STO (b) compared to thin ($t$ = 17 nm) films on STO (c) and Ti-STO (d). The maps are recorded in the vicinity of the (204) STO reflection. The intensity is plotted on a logarithmic scale as a function of the scattering vector $q$ expressed in noninteger Miller indices $h$, $k$, and $l$ of the STO substrate reflection, referring to the [001] surface normal and the azimuth references [100], [010], [-100], and [0-10]. The SIO film peaks (444), (260), (44-4), and (620) are assigned on the basis of the orthorhombic notation and correspond to the {204} pseudo-cubic lattice reflections alike. In-plane and out-of plane reciprocal lattice spacing of the substrate are indicated by dotted lines. Dashed lines are guide to the eye to visualize different film peak positions. Reflections of different domains are indicated in different colors.

The growth on Ti-STO (Fig. 2b) results in twinning alike. However, peak position is less shifted to smaller $h$-values and indicates a somewhat larger out-of-plane lattice spacing. The residual strain might be favored by the lower amount of surface irregularities of Ti-STO. For $t < 30$ nm, films usually show strained growth. In Figs. 2c,d we plot the RSMs for SIO on STO and Ti-STO for $t = 17$ nm. The reduced film thickness results in lower peak intensities. All the peaks are positioned at $|h|= 2$, indicating coherent film strain with projected in-plane lattice spacing identical to that of STO. In contrast, intensity along the $l$-direction appears to be broadened. For SIO on STO the $l$-values for the two orthogonal azimuth directions [100] and [010] are the same and slightly smaller compared to those of the [-100] and [0-10]. The peak pattern cannot be related to a single symmetry group but strongly suggests twinned growth of SIO. The RSM and epitaxial relationship of the two SIO domains (indicated by green and blue color) is sketched in Figs. 3a,b. For a given azimuth, peak intensities of the two domains overlap resulting in a distinct broadening along the $l$-direction. The untwinned growth, documented in Fig. 2d, indicates a single domained growth with [-110] SIO parallel to [010] STO, i. e., in comparison to the green symbols and schematic in Figs. 3a,b, a domain rotated by 180°. Obviously, the orthorhombic $a$- and $b$-lattice parameters of SIO are different whereas the in-plane lattice spacing appears to be identical. In addition, ω–scans reveal that [110] SIO is parallel to [001] STO which indicates the presence of a small monoclinic distortion, see Fig. 3c. From the lattice spacing a monoclinic angle $\gamma \approx 88°$ is deduced, which has been reported also by others for thin SIO films on STO [25]. Strictly speaking, for orthorhombic symmetry $\alpha = \beta = \gamma = 90°$. Nevertheless, since monoclinic distortion ($\gamma < 90°$) for SIO on STO and Ti-STO is rather small we kept the orthorhombic notation for the SIO film throughout. For SIO on Ti-STO, the peak pattern (see Fig. 2d) indicates the growth of single domain, i. e., untwinned orthorhombic SIO. The $l$-values for the (444) and (44-4) film peaks are the same, whereas those for the (620) and (260) are distinct smaller and larger, respectively. A monoclinic distortion is verified alike and comparable to that found for SIO films with similar film thickness on STO. Within the experimental accuracy, the refined lattice parameters of the thin SIO (17 nm) films on STO and Ti-STO are $a = (5.6 \pm 0.01)$Å, $b = (5.64 \pm 0.01)$Å, and $c = (7.81 \pm 0.02)$Å, where $\gamma = 88° \pm 0.2°$. For thick relaxed SIO (60nm) we deduce $a = (5.58 \pm 0.02)$Å, $b = (5.58 \pm 0.02)$Å, and $c = (7.82 \pm 0.04)$Å, where $\gamma = 88.8° \pm 0.3°$. Compressive in-plane strain obviously promotes monoclinic distortion.

Interestingly, the $c$-axis of SIO displays alignment parallel to the step-edges, i. e., the (444) and (44-4) peaks appear for [100] and [-100] azimuth direction, respectively.

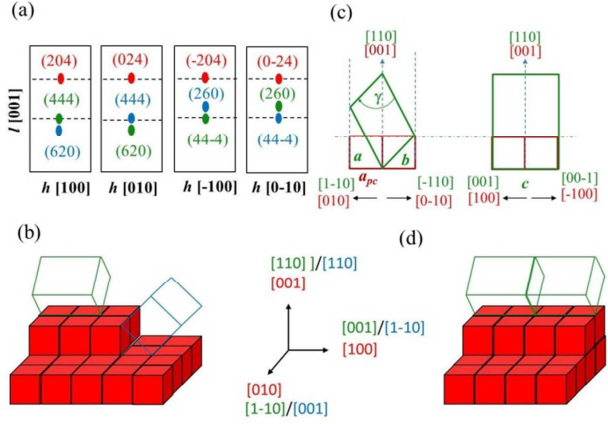

Figure 3. (a) Scheme of a reciprocal space map of a twinned SIO film on STO as shown in (b). The STO substrate peaks are indicated in red. Film peak intensity from two orthogonal domains (shown in blue and green) overlap and may appear as one peak in the diffraction experiment with distinct broadening along the $l$-direction, see Fig. 2c. (b) Schematic of two orthogonal domains (green and blue) grown on standard STO (red). Crystallographic directions are indicated. (c) The RSM shown in (a) demonstrates orthorhombic lattice parameters $a \neq b$ for SIO which, in combination with identical in-plane lattice spacing along all four azimuth directions, strongly suggests a monoclinic distortion between the $a$- and $b$-axis with $\gamma < 90°$. (d) The deposition of SIO on Ti-STO results in an untwinned, single domained growth of SIO with $c$-axis orientation parallel to the step-edges of substrate.

These results were reproduced on several samples and seem to be very robust. For step-edge alignment distinct different from [100] or [010] STO direction, untwinned growth diminishes. More quantitative studies, not done yet, may give the possibility to tune systematically the degree of twinning. In contrast, films on DSO display single domained orthorhombic growth independent of $t$ with $c$-axis orientation parallel to that of DSO [23,25].

In Fig. 4a we have plotted the resistivity of non-patterned SIO films ($t = 17$ nm) on STO and Ti-STO for the [100]- and [010]-substrate directions. For the twinned films on STO (see Fig. 3c), resistivity is isotropic and slightly increases with decreasing $T$. Consistent with previous results, the normalized resistivity ratio $r_n = [\rho_{010}(T)/\rho_{010}(300K)]/[\rho_{100}(T)/\rho_{100}(300K)] \approx 1$, see Fig. 4b. Here, $\rho_{010}(T)$ and $\rho_{100}(T)$ is the $T$-dependent resistivity along the [010] and [100]-direction of STO. In stark contrast, the resistivity of the untwinned film on Ti-STO displays strong anisotropic electronic transport with $\rho_{010} > \rho_{100}$, i. e., with respect to the orthorhombic axes of SIO: $\rho_{1-10} > \rho_{001}$. $r_n$ steadily increases with decreasing $T$ exceeding 1.6 below 20 K. In Fig. 4b, we likewise included $r_n$ of bulk-like SIO on DSO for which $r_n$ first shows similar behavior with decreasing $T$. However, below 200 K, $r_n$ shows a distinct $T$-dependence with a maximum of about 1.3 at $T \approx 85$ K. The anomalous thermal expansion of orthorhombic lattice parameters of bulk SIO results in a distinct $T$-dependence of the structural in-plane anisotropy between the [1-10] and [001] SIO direction [19], i. e., $(a^2+b^2)^{1/2}/c(T)$ which is very similar to that of the electrical anisotropy $\rho_{1-10}/\rho_{001}$ ($r_n(T)$). For that reason, the electrical anisotropy is very likely related to the anomalous thermal expansion of SIO and intrinsic, dominated by octahedral distortions. Obviously, the electronic anisotropy ($r_n-1$) of untwinned and strained SIO films on Ti-STO exceeds intrinsic anisotropy by a factor of two.

A more detailed study on anisotropic transport is shown in Figs. 4c,d, where FPP measurements on microbridges are shown for SIO on Ti-STO and for comparison on DSO for various $t$. For untwinned SIO on Ti-STO $r_n$ increases with increasing $t$ for $t \leq 27$ nm. The $t$-dependence of $r_n$ is very likely explained in terms of substrate-induced clamping which impedes SIO-like thermal expansion or octahedral distortion. In addition, for the thinnest film ($t = 2.9$ nm) a transition towards tetragonal structure may evolve [24] which further reduces $r_n$. Films with $t = 50$ nm display twinned growth and therefore $r_n \approx 1$ again. Similar to the VDP measurements on non-patterned films, $r_n$ steadily increases with decreasing $T$, however, with smaller amplitude. In contrast, patterned films on DSO show no specific $t$-dependence for $t > 2.5$ nm due to the much smaller lattice strain and a thermal expansion comparable to that of DSO [19,26]. Therefore, the distinct $T$-dependence of $r_n$ is less affected. Deviations for $t = 2.9$ nm are likely caused by residual strain. Again, the electronic anisotropy is roughly only half of that of patterned films on Ti-STO and reduced with respect to plane films alike. The reduced amplitude of $r_n$ with respect to plane films results from a somewhat larger resistivity of the microbridges at low $T$. This might be caused by the limited applicability of the van der Pauw technique to electrical homogeneous systems or aging effects during Ar-ion milling which may increase resistivity of patterned films alike.

The experiments demonstrate that twinning of SIO can be suppressed by epitaxial growth on Ti-STO. We assume, that the suppressed growth of other domains is strongly related to the growth kinetics of SIO on Ti-STO and STO. The untwinned growth of SIO on Ti-STO allowed us to study electronic anisotropy in more detail. In comparison to bulk-like untwinned SIO strained films on Ti-STO ($t < 30$ nm) display significant larger electronic anisotropy at low $T$. The increase of $r_n$ is most probably related to compressive strain. The reduced $c$-lattice parameter likely results in a shortening of the Ir-O-bond length and hence increase of the Ir5$d$-O2$p$ orbital hybridization which again decreases $\rho_{001}$, whereas additional monoclinic distortion reduces Ir-O-Ir bond angle and therefore hybridization along the [1-10] direction leading to an increase of $\rho_{1-10}$. Surface engineering may give the possibility of tuning the degree of twinning and therefore anisotropic electronic transport in SIO.

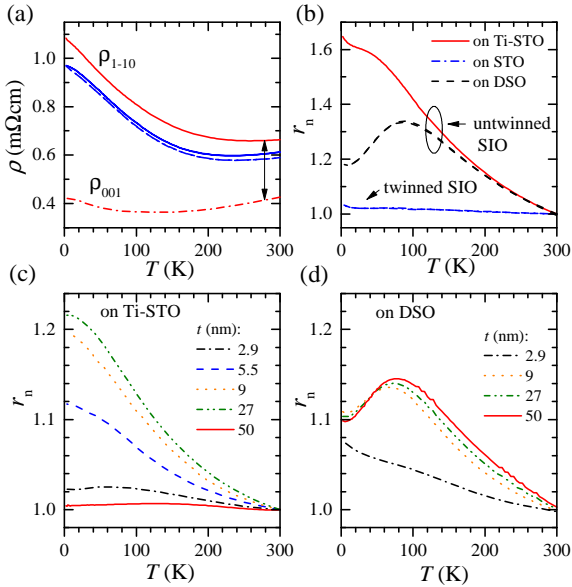

Figure 4. (a) $\rho$ versus $T$ for current flow parallel to the [010]- (solid line) and [100]-(dashed-dotted line) substrate direction for non-patterned SIO films (17 nm) on STO (blue) and Ti-STO (red). For untwinned SIO on Ti-STO the orthorhombic directions are indicated. (b) Resulting normalized resistivity ratio $r_n = [\rho_{010}(T)/\rho_{010}(300K)]/[\rho_{100}(T)/\rho_{100}(300K)]$ (see text) versus $T$. For comparison, $r_n$ of bulk-like SIO (50 nm) on DSO is shown. $r_n$ versus $T$ determined by measurements on microbridges for SIO on Ti-STO (c) and DSO (d) for various film thickness $t$.

Acknowledgement

We are grateful to R. Thelen and the Karlsruhe Nano Micro Facility (KNMF) for technical support and J. Schubert from the Peter Grünberg Institut, Forschungzentrum Jülich, for providing sample analysis by Rutherford backscattering spectrometry. D.F. also acknowledges K. Sen and S. Mukherjee for fruitful discussion.